\documentstyle[aps,prabib]{revtex}
\onecolumn
\tighten
\begin{document}
\title{Low-energy interaction of composite spin-half systems with
scalar and vector fields}

\author {D.~R. Phillips${}^1$, M.~C. Birse${}^{1,2}$, and S.~J. Wallace${}^1$}

\address{${}^1$ Department of Physics and Center for Theoretical Physics,\\
University of Maryland, College Park, MD, 20742-4111, USA\\
${}^2$ Theoretical Physics Group, Department of Physics and Astronomy,\\
University of Manchester, M13 9PL, UK}

\date{\today}
\maketitle

\begin{abstract}
We consider a composite spin-half particle moving in spatially-varying
scalar and vector fields. The vector field is assumed to couple to a
conserved charge, but no assumption is made about either the structure
of the composite or its coupling to the scalar field.  A general form
for the piece of the spin-orbit interaction of the composite with the
scalar and vector fields which is first-order in momentum transfer
${\bf Q}$ and second-order in the fields is derived.
\end{abstract}

\section{Introduction}

In the low-energy limit, the interaction of a composite spin-half
system, such as a nucleon, with external scalar and vector fields has
been shown to be governed, at least in part, by general principles of Lorentz
invariance~\cite{Bi95}.  Consideration of a nucleon in a
spatially-uniform, scalar field provides the simplest result: the only
effect is that the mass becomes a function of the strength, $S$, of
the scalar field, {\it i.e.} $M \rightarrow M(S)$.  If a vector
interaction coupled to a conserved charge is also present, then the
energy of the composite particle may be expressed as
\begin{equation}
E = \sqrt{ M^2(S) + {\bf p} ^2 } ~ + ~ V,
\label{eq:EofMofS}
\end{equation}
where ${\bf p}$ is the three-momentum of the particle and $V$ is the
strength of the vector interaction.

Naturally, Eq.~(\ref{eq:EofMofS}) may be derived using analyses based
on quantum field theory. Following such an approach,
Refs.~\cite{Wa94,Wa95} considered a specific model of a spin-half
composite. In that model the composite interacts with an external
scalar field which provides a pure shift of the mass, {\it i.e.}
$M(S)=M+S$. A vector field is also introduced, and couples to the
(conserved) baryonic charge of the model. This leads to the following
expansion for the energy of the composite particle,
\begin{equation}
E = \epsilon ({\bf p}) + \frac{M}{\epsilon ({\bf p})} S + V + 
\frac{{\bf p}^2 S^2}{2 \epsilon ^3 ({\bf p})} + \cdots ,
\label{eq:EofSexp}
\end{equation}
where $\epsilon ({\bf p}) \equiv \sqrt{M^2 + {\bf p}^2}$ is the
relativistic energy when the scalar field is zero.  Of course, these
are the first few terms in the expansion of Eq.~(\ref{eq:EofMofS}) for
the case of a simple mass shift.  Since specific choices for the model of
compositeness and the coupling of the scalar field to the
composite were made in Refs. \cite{Wa94,Wa95} the derivation of
Eq.~(\ref{eq:EofSexp}) is less general than that of Ref.~\cite{Bi95}.

These results provide important confirmation of effects which were
discovered some time ago to be crucial ingredients in successful
descriptions of spin observables in the elastic scattering of protons by
nuclei at intermediate energy.  In a relativistic description, the
nucleon-nucleon interaction is known to possess a strong scalar
attraction and a strong vector repulsion~\cite{Wa87}.  This translates
into optical potentials which have the same character for nucleon-nucleus
scattering.  For nucleon-nucleus interactions evaluated at nuclear matter
density, the terms in Eq.~(\ref{eq:EofSexp}) which are linear in $S$ and $V$
end up largely canceling, with the net interaction being about 50 MeV of
attraction.  The term that is quadratic in $S$ then becomes significant 
because of the comparatively large (roughly 300 MeV) scalar attraction.

Relativistic descriptions of nucleon-nucleus scattering employ the Dirac
equation with scalar and vector potentials and thus include the
term in Eq.~(\ref{eq:EofSexp}) which is quadratic in $S$. In such an
approach this term arises from z-graphs. Therefore analyses based
upon the Schr\"{o}dinger equation with relativistic kinematics omit
it. This omission has been justified by arguing that the excitation of
virtual $N \bar{N}$ pairs should be suppressed due to compositeness of
the nucleon~\cite{Br84,JB91,Co92}.  Recent
works~\cite{Bi95,Wa94,Wa95} show that the appearance of the term which
is quadratic in $S$ in Eq.~(\ref{eq:EofSexp}) is a consequence of
Lorentz invariance. Thus, this term will {\it not} be suppressed by
compositeness in the low-energy limit.

While the analysis of Ref.~\cite{Bi95} makes no assumptions about the
composite structure or the dynamics governing its interaction with the
scalar field, the quantum field theory approach of
Refs.~\cite{Wa94,Wa95} is limited to a simple model of compositeness
and a scalar field which couples to the divergence of the dilatation
current of the composite model.  Furthermore, both analyses apply only
to spatially-uniform fields. We wish to remove these limitations by
performing a quantum-field-theoretic analysis of this problem which is
model independent. Such an approach allows us to explore whether
there is a model-independent prediction for the spin-orbit interaction
of the composite with the scalar and vector fields. 

The first-order contribution to the spin-orbit force is a 
consequence of Lorentz invariance and gauge invariance (see, for 
example, Ref.~\cite{AW88}). Simple
arguments (see Sec.~\ref{sec-five} for details) show that the
first-order spin-orbit interaction for the simple mass-shift case is,
to first-order in ${\bf Q}$:
\begin{equation}
V^{(1)}_{SO}({\bf p}+ {\bf Q}/2,{\bf p}-{\bf Q}/2)=
\left(S - V - 
V \kappa \frac{\epsilon({\bf p}) + M}{2 M}\right)
\frac{i}{2 \epsilon({\bf p})}\frac{1}{\epsilon({\bf p}) + M} {\bf \sigma}
\cdot ({\bf p} \times {\bf Q}),
\label{eq:VSO1}
\end{equation}
where $\kappa$ is the anomalous magnetic moment of the composite
system under consideration. The opposite sign of the vector
contribution in Eqs.~(\ref{eq:EofSexp}) and (\ref{eq:VSO1}) means that
if the scalar and vector potentials $S$ and $V$ are large and
approximately equal in magnitude but opposite in sign, then the central
piece of the potential will be small and the spin-orbit force
large. This ability to achieve cancellation in the central force with
reinforcement in the spin-orbit force is one of the key ingredients of
the success of Dirac phenomenology~\cite{Wa87}.

However, Eq.~(\ref{eq:VSO1}) tells us nothing about the piece of
$V_{SO}$ which is higher-order in the fields. In the case that the
scalar interaction is zero we know that there is a model-independent
result (provided the external vector field couples to a conserved
charge) because of the low-energy theorem for Compton
scattering~\cite{Lo54,GG54}. The question addressed in this work is
whether a similar model-independent result exists if both scalar and
vector fields are present.

To this end we present an analysis, based on quantum
field theory, of the behavior of a spin-half composite in scalar and
vector fields. The scalar interaction is arbitrary, while the vector
field is assumed to couple to a conserved charge.  The analysis
retains terms which contribute to the spin-orbit interaction, namely
those that are first-order in the spin and momentum transfer. The
object is to determine the part of the spin-orbit interaction that is
second-order in the strengths of the external interactions. This
part comes mainly from z-graphs if the Dirac equation is used.  The
results extend the earlier analyses by showing that a composite system
interacts at low momentum transfer in a similar fashion to a Dirac
particle whose mass and anomalous magnetic moment are functions of the
external scalar field.

The analysis proceeds as follows. In Sec.~\ref{sec-two} we write
down the composite particle Green's function, define its couplings to
scalar and vector fields, and write down the amplitude for the
composite's second-order interaction with these fields. In
Sec.~\ref{sec-three} we show that if one is interested only in
terms to first-order in the momentum transfer then the amplitude may
be evaluated with one interaction carrying the total momentum transfer,
${\bf Q}$, and the other carrying zero momentum.
Section~\ref{sec-four} then derives a particularly simple relation for
this amplitude using identities for the Green's function in constant
fields.  This result relates the irreducible second-order interaction
with one insertion carrying zero momentum to a derivative of the
first-order interaction in an arbitrary scalar field.  In
Sec.~\ref{sec-five} we write the general forms of the first-order
scalar and vector vertices, correct to first-order in ${\bf Q}$, then
use the result of Sec.~\ref{sec-four} to calculate the spin-orbit
piece of the second-order interaction, also correct to first-order in
${\bf Q}$.  Conclusions are drawn in Sec.~\ref{sec-conc}.

\section{The second-order scattering amplitude}

\label{sec-two}

Consider a generic model field theory which leads to a bound spin-half state
with microscopic dynamics described by the Lagrangian, ${\cal L}_0$.
Suppose that we construct an interpolating field for the spin-half composite,
$\psi(x)$, out of the microscopic degrees of freedom of the theory. The field
$\psi$ may be chosen to be any combination of constituent field operators with
the appropriate quantum numbers. In particular, we may assume, without loss of
generality, that $\psi$ transforms as a Dirac spinor under Lorentz
transformations. One example of an interpolating field with this property is
the Ioffe field commonly used in QCD sum rule calculations~\cite{Io81}. We
then examine the Green's function,
\begin{equation}
G(p) \equiv
i\int d^4(x' - x) e^{i p (x'-x)} \langle 0|T(\psi(x') \overline{\psi}(x))|0
\rangle.
\label{eq:Gdef}
\end{equation}
The operator $\overline{\psi}=\psi^\dagger \gamma_0$ since $\psi$
transforms as a Dirac spinor under Lorentz transformations.
If the composite has mass $M$ then we know that this Green's function
takes the form~\cite{IZ84}
\begin{equation}
G(p)=\frac{Z_2 u({\bf p}) \bar{u}({\bf p})}
{p^0 - \epsilon({\bf p}) + i \eta} + \delta g(p),
\label{eq:G}
\end{equation}
where $Z_2$ is the so-called wave function renormalization and $\delta g$
is the piece of the Green's function which is regular at the pole
$p_0=\sqrt{M^2 + {\bf p}^2}$. The ``wave function'' $u({\bf p})$ is defined by
\begin{equation}
Z_2^{1/2} u({\bf p}) e^{-ipx}=\langle 0|\psi(x)|N({\bf p}) \rangle,
\label{eq:udef}
\end{equation}
where $p_0=\sqrt{M^2 + {\bf p}^2}$, and $|N({\bf p}) \rangle$ is a
Fock space state representing a composite with momentum ${\bf p}$ and
mass $M$. From (\ref{eq:udef}) we see that $\bar{u}=u^\dagger
\gamma_0$.  Since the operator $\psi$ transforms as a Dirac spinor
under Lorentz transformations, $u({\bf p})$ must transform as a Dirac
spinor corresponding to a particle of mass $M$ under such
transformations. Therefore, up to a normalization factor, $u({\bf p})$
{\it is} a Dirac spinor for a particle of mass $M$.  The constant
$Z_2$ is then chosen so that $u^\dagger ({\bf p}) u({\bf p})=1$.

Now let us couple scalar and vector fields to this composite. In this work we
do not specify how the scalar field, $\sigma$,
couples to the microscopic constituents. By
contrast, the vector field, $\omega$, is introduced by minimal
substitution, {\it i.e.} by making the replacement
\begin{equation}
i \partial_0 \rightarrow i \partial_0 - \omega(x).
\label{eq:minsub}
\end{equation}
The vector field therefore couples to a conserved charge. If this
vector field is constant, its interaction Hamiltonian commutes with
the Hamiltonian describing the internal dynamics of the composite, and
thus cannot produce a change in the structure of the composite; it can
only shift its energy. This also implies that there is no change in the
wavefunction renormalization in the presence of a uniform $\omega$
field.

The total Lagrangian, ${\cal L}$, of this theory may be expressed as, 
\begin{equation}
{\cal L}(x)={\cal L}_0(x) - \rho_0(x) \omega(x) - \rho^{(1)}(x) \sigma(x)
- \rho^{(2)}(x) \sigma^2(x) - \ldots,
\label{eq:lag}
\end{equation}
where $\rho_0(x)$ is the vector charge density and $\rho^{(n)}(x)$ is
the scalar density through which the nth power of the scalar field
couples.  In the analysis of Refs.~\cite{Wa94,Wa95}, the $\sigma$ field
(which in that case is uniform) couples to the divergence of the
dilatation current of the theory, and therefore $Z_2$ can only depend
on powers of the scalar field.  Since $Z_2$ is a dimensionless
quantity, it follows that in the model of Refs.~\cite{Wa94,Wa95} it
has no dependence on the scalar field whatsoever.  However, general scalar 
couplings do not lead to this simple scaling behavior and, even if 
the scalar field is spatially uniform, it can change the internal structure of
the composite fermion in a nontrivial way. Hence, in general, the
wavefunction renormalization, $Z_2$, will be a function of the scalar field.

Having defined the type of model and given the form that the Green's
function takes in the absence of any scalar and vector fields, we next
write down the Green's functions and amplitudes for the interactions of
the composite with these external fields.  In general, the Green's
function for the composite moving in these fields may be written as a
functional integral:
\begin{equation}
G(x',x;\sigma,\omega)=\int [d \zeta] \psi(x')
 \overline{\psi}(x) \exp(i {\cal S}[\zeta,\sigma,\omega]),
\label{eq:funct}
\end{equation}
where it is understood that the arguments after the semicolon on
the left-hand side are functions, while those before it are variables.
Here $\zeta$ represents all fields present in the
vacuum composite Lagrangian, ${\cal L}_0$, and 
${\cal S}[\zeta,\sigma,\omega]$ is the action
corresponding to the Lagrangian
defined in Eq.~(\ref{eq:lag}). In order
to obtain the interaction of the composite with the fields at
a given order, the Green's function (\ref{eq:funct}) must be expanded
in powers of the external fields $\omega$ and $\sigma$. 

We now assume that external scalar and vector fields are present with 
the same spatial distribution function, $h({\bf r})$. This assumption is 
made for simplicity, and is not essential to our argument. The function
$h({\bf r})$ is normalized so that $h({\bf 0})=1$. The scalar and vector 
distributions are taken to have the forms,
\begin{equation}
\sigma ({\bf r})=S~h({\bf r}); \quad \omega ({\bf r})=V~h({\bf r}),
\end{equation}
where $S$ and $V$ are strength parameters for the scalar and vector
interactions.  

The first-order amplitude for the interaction of the composite with
these fields may be written as $h({\bf Q})A^{(1)}(p',p,Q)$, where 
$h({\bf Q})$ is the Fourier transform of
$h({\bf r})$, and the function $A^{(1)}(p',p,Q)$ is defined via
\begin{equation}
G(p') A^{(1)}(p',p,Q) G(p)
\equiv \int d^4x' \, d^4x \, d^4y_1 \, e^{i (p' x'-p x - Q y_1)} \left.
D_{y_1} G(x',x;\sigma,\omega) \right|_{\sigma=\omega=0},
\label{eq:firstorder}
\end{equation}
where
\begin{equation}
D_y=S \frac{\delta}{\delta \sigma(y)} + V \frac{\delta}{\delta \omega(y)}.
\end{equation}
For convenience, the strengths of the scalar and vector interactions are 
included in the vertex function.
This allows us to define a first-order vertex function, $\Lambda$, 
which is a combination of scalar and vector vertex functions:
\begin{equation}
A^{(1)}(p',p,Q) \equiv (2 \pi)^4 \delta^{(4)}(p'-p-Q) \Lambda(p',p).
\label{eq:lambdadef}
\end{equation}
The first-order potential is then found by multiplying
$\Lambda(p',p)$, with $p$ and $p'$ both on-shell, by $h({\bf Q})$. 
If $h({\bf r})$ is constant this
implies that $\Lambda(p',p)$ will be needed only at $p'=p$, {\it i.e.}
$Q=0$.  From Eqs.~(\ref{eq:firstorder}) and (\ref{eq:lambdadef}) we
see that the $Q = 0$ vertex function is obtained from
\begin{equation}
G(p) \Lambda(p,p) G(p)=\int d^4(x'-x) e^{i p (x'-x)}
\int d^4y_1 \left. D_{y_1} G(x',x;\sigma,\omega) \right|_{\sigma=\omega=0}.
\label{eq:zeropvert}
\end{equation}
In general, if $F$ is a functional of the function $g$, then
\begin{equation}
\int d^4y' \, \left.\frac{\delta}{\delta g(y')}  F[g] \right|_{g=0}=\left.
\frac{\partial F(g)}{\partial g} \right|_{g=0},
\label{eq:functid}
\end{equation}
where on the right-hand side, $F$ is to be thought of as a {\it function} of
the constant parameter $g$.  Thus the $Q = 0$ vertex function,
corresponding to spatially-uniform external fields, is given by
\begin{eqnarray}
G(p) \Lambda(p,p) G(p)&=&D \left. G_{\sigma,\omega}(p)
\right|_{\sigma=\omega=0},
\label{eq:Lamfunct1}\\
D&=&S \frac{\delta}{\delta \sigma} + V \frac{\delta}{\delta \omega},
\label{eq:Lamfunct2}
\end{eqnarray}
where $G^{\sigma,\omega}$ is calculated via Eq.~(\ref{eq:funct}), but
with constant $\sigma$ and
$\omega$ fields. Moreover, because the $\omega$ field is introduced by minimal
substitution Eqs.~(\ref{eq:Lamfunct1}) and (\ref{eq:Lamfunct2}) may be 
rewritten as
\begin{eqnarray}
G(p) \Lambda(p,p) G(p)&=&D \left. G_{\sigma}(p)\right|_{\sigma=0},
\label{eq:vertid}\\
D&=&S \frac{\delta}{\delta \sigma} - V \frac{\delta}{\delta p_0},
\end{eqnarray}
and $G_\sigma(p)$ is the Green's function calculated at arbitrary constant
scalar field $\sigma$. This result will be used in Sec.~\ref{sec-four} 
as the basis for a straightforward calculation of the first-order 
vertex function.

In order to calculate the second-order scattering of the composite in
scalar and vector fields we define
the amplitude $A^{(2)}$ via
\begin{equation}
G(p') A^{(2)}(p',p,q_2,q_1) G(p)
\equiv {1 \over 2} \int d^4x' \, d^4x \, d^4y_2 \, d^4 y_1 \, 
e^{i (p'x' - px - q_2 y_2
- q_1 y_1)} \left. D_{y_2} D_{y_1} G(x',x;\sigma,\omega) 
\right|_{\sigma=\omega=0}.
\label{eq:A2def}
\end{equation}
If $A^{(1)}(p',p,Q;\sigma,\omega)$ is defined exactly as
in Eq.~(\ref{eq:firstorder}), but with all functions retaining their 
dependence on the fields $\sigma$ and $\omega$, i.e.,
\begin{equation}
G(p';\sigma,\omega) A^{(1)}(p',p,Q;\sigma,\omega) G(p;\sigma,\omega)
\equiv \int d^4x' \, d^4x \, d^4y_1 \, e^{i (p' x'-p x - Q y_1)}
D_{y_1} G(x',x;\sigma,\omega),
\end{equation}
then Eq.~(\ref{eq:A2def}) may be rewritten as 
\begin{equation}
G(p') A^{(2)}(p',p,q_2,q_1) G(p)={1 \over 2} \int d^4 y_2 \, e^{-i q_2 y_2} 
\left. 
D_{y_2} \bigl[G(p';\sigma,\omega)
A^{(1)}(p',p,q_1;\sigma,\omega) G(p;\sigma,\omega)\bigr] 
\right|_{\sigma=\omega=0}.
\end{equation}
Next, observe that a contact term, corresponding to interaction within
the vertex function $A^{(1)}$ and denoted by $C$, may be defined via
\begin{equation}
\int d^4y_2 \, e^{-i q_2 y_2} \left. D_{y_2} A^{(1)}(p',p,q_1;\sigma,\omega)
\right|_{\sigma=\omega=0}=(2 \pi)^4 \delta^{(4)}(p'-p-q_1-q_2) 
C(p+q_1+q_2,q_2,q_1,p).
\label{eq:contactdef}
\end{equation}
Using Eqs.~(\ref{eq:contactdef}), (\ref{eq:firstorder}), and 
(\ref{eq:G}) in Eq.~(\ref{eq:A2def}), and then taking the limit as
the initial and final state poles are approached, 
yields the following physical 
amplitude for the second-order interaction of the 
composite with the fields at total three-momentum transfer ${\bf Q}$:
\begin{eqnarray}
&& \bar{u} ({\bf p} + {\bf Q}/2) {\cal
A}(\overline{p+Q/2},Q/2-q, Q/2+q,\overline{p-Q/2}) u({\bf p} - {\bf Q}/2) 
\equiv \lim_{p_0 + Q_0/2 \rightarrow
\epsilon({\bf p} + {\bf Q}/2)} \, \, \lim_{p_0 - Q_0/2 \rightarrow
\epsilon({\bf p} - {\bf Q}/2)}\nonumber\\ 
&& \qquad 
\left\{{1 \over 2} Z_2 \bar{u} ({\bf p} + {\bf Q}/2) 
\Lambda (p + Q/2,p+q) G(p+q) \Lambda(p+q,p-Q/2) u({\bf p}
- {\bf Q}/2) \right. \nonumber\\ 
&& \qquad \qquad + {1 \over 2} Z_2 \bar{u} ({\bf p} + {\bf Q}/2)
\Lambda (p + Q/2,p-q) G(p-q) \Lambda(p-q,p-Q/2)
u({\bf p} - {\bf Q}/2) \nonumber\\ 
&& \left. \qquad \qquad \qquad + {1 \over 2} Z_2 \bar{u} ({\bf
p} + {\bf Q}/2) C(p+Q/2,Q/2-q,Q/2+q,p-Q/2) u({\bf p}
- {\bf Q}/2) \right\}.
\label{eq:medef}
\end{eqnarray}
Here we have made use of the overall delta function and then defined a new
amplitude ${\cal A}$ which does not include it.
Note that the initial and final momenta both depend on $Q$, and the bars
on the initial and final arguments of ${\cal A}$
indicate that both external legs are on-shell, {\it i.e.} $\overline{k}=
(\epsilon({\bf k}),{\bf k})$.
This determines $Q_0$, and implies that $p^2=M^2$ to first-order 
in $Q$. Finally, in order to generate a potential, the 
matrix element (\ref{eq:medef}) 
must be integrated against the field distribution, $h$:
\begin{eqnarray}
&& {\cal V}^{(2)}({\bf p}+{\bf Q}/2,{\bf p}-{\bf Q}/2)\nonumber\\ 
&& \qquad = \int
\frac{d^3q}{(2 \pi)^3} \, h({\bf Q}/2+{\bf q}) h({\bf Q}/2-{\bf q}) Z_2 
\bar{u} ({\bf p} + {\bf Q}/2) {\cal
A}(\overline{p+Q/2},Q/2-q, Q/2+q,\overline{p-Q/2}) u({\bf p} - {\bf Q}/2).
\label{eq:Vdef}
\end{eqnarray}

\section{Low-energy limit of the second-order amplitude}

\label{sec-three}

So far no approximations have been made, and the matrix element
(\ref{eq:medef}) could (in principle) be obtained by evaluating the
relevant Feynman diagrams for a given composite model Lagrangian
${\cal L}_0$, as was done in Refs.~\cite{Wa94,Wa95} for the case
${\bf Q}=0$. However, a result which is independent of the details 
of both the sub-structure of the composite and its coupling to the
scalar field may be obtained by taking the low-energy limit of this
expression.  In particular, in this section we show that in the limit
of slowly-varying external fields we may restrict our analysis to the
case where one of the external field interactions carries all the
momentum and the other occurs at zero momentum.

First, note that the amplitude ${\cal A} $ corresponds to a sum of
direct and crossed Feynman diagrams and therefore is symmetric with
respect to interchange of the momenta of the external fields.
Defining $p_f=p+q_1+q_2$ and $p_i=p$, this implies that
\begin{equation}
{\cal A} (p_f,q_1,q_2,p_i)={\cal A}(p_f,q_2,q_1,p_i),
\label{eq:crossing}
\end{equation}
and it follows that
\begin{equation}
\left. \frac{\partial {\cal A}(p_f,q_1,q_2,p_i)}{\partial q_1^\mu}
\right|_{q_1=q_2}=
\left. \frac{\partial {\cal A}(p_f,q_1,q_2,p_i)}{\partial q_2^\mu}
\right|_{q_1=q_2}.
\label{eq:pdsymm}
\end{equation}

Expanding the amplitude ${\cal A}(p_f,Q/2-q,Q/2+q,p_i)$ 
in powers of $Q/2+q$ and $Q/2-q$ and using Eq.~(\ref{eq:pdsymm}) produces
\begin{equation}
{\cal A}(p_f,Q/2-q,Q/2+q,p_i)={\cal A}(p_f,0,0,p_i)
+ Q^\mu \left. \frac{\partial {{\cal A}(p_f,q_1,0,p_i)}}{\partial q_1^\mu}
\right|_{q_1=0} + {\cal O}[(Q/2+q)^2] + {\cal O}[(Q/2-q)^2].
\label{eq:identity1}
\end{equation}
Hence, if we are interested only in terms of order $Q$, we may write
\begin{eqnarray}
{\cal A}(p_f,Q/2-q,Q/2+q,p_i)&=&{\cal A}(p_f,Q,0,p_i) + 
{\cal O}[(Q/2+q)^2] + {\cal O}[(Q/2-q)^2]\nonumber\\
&=&{\cal A}(p_f,0,Q,p_i) + {\cal O}[(Q/2+q)^2] + {\cal O}[(Q/2-q)^2],
\end{eqnarray}
where the last equality follows from Eq.~(\ref{eq:crossing}).

When this equation is inserted into the definition (\ref{eq:Vdef}), the 
terms of second and higher order in momentum transfer generate second and 
higher derivatives of the field distribution, $h({\bf r})$. Provided that
$h$ is slowly varying in space, these may be neglected and we obtain (correct
to first-order in ${\bf Q}$):
\begin{equation}
{\cal V}^{(2)}({\bf p}_f,{\bf p}_i) =\frac{1}{2} \int \frac{d^3q}{(2
\pi)^3} \, h({\bf Q}/2+{\bf q}) h({\bf Q}/2-{\bf q}) Z_2 \bar{u} ({\bf
p}_f) {\cal A}(\overline{p_f},Q,0,\overline{p_i}) u({\bf p}_i).
\label{eq:potform}
\end{equation}

\section{A general result for the amplitude with one
zero-momentum insertion}

\label{sec-four}

Our next task is to evaluate the matrix element
$\bar{u} ({\bf p}_f) {\cal A}(\overline{p_f},Q,0,\overline{p_i}) u({\bf p}_i)$.
From (\ref{eq:medef}),
\begin{eqnarray}
{\cal A}(\overline{p_f},Q,0,\overline{p_i})&=&
\lim_{p_i^0 \rightarrow \epsilon({\bf p}_i)}
{1 \over 2} Z_2 \Lambda (\overline{p_f},p_i) G(p_i)
\Lambda(p_i,\overline{p_i})\nonumber\\
&+& \lim_{p_f^0 \rightarrow \epsilon({\bf p}_f)}
 {1 \over 2} Z_2 \Lambda (\overline{p_f},p_f) G(p_f) 
\Lambda(p_f,\overline{p_i}) + {1 \over 2} Z_2 C(\overline{p_f},Q,0,
\overline{p_i}),
\label{eq:intamp}
\end{eqnarray}
where a bar over a four-vector indicates that it
is constrained to its mass shell.  The limits here are equivalent to
taking the limit $q \rightarrow  -{1 \over 2} Q$ in Eq. (\ref{eq:medef}).
We wish to relate the vertices which 
appear in this equation to zero momentum-transfer vertices, since we know
that these may be obtained using Eq.~(\ref{eq:vertid}).

Indeed, before we begin our effort to calculate 
$\bar{u}({\bf p}_f) {\cal A}(p_f,Q,0,p_i) u({\bf p}_i)$,  
we derive two sets of identities from Eq.~(\ref{eq:vertid}).
First, note that the Green's function $G_\sigma$ may be written as
\begin{equation}
G_\sigma (p)=\frac{Z_2(\sigma) u_\sigma({\bf p}) \bar{u}_\sigma ({\bf p})}
{p^0 - \epsilon_\sigma({\bf p})} + \delta g_\sigma(p),
\label{eq:GFsigma}
\end{equation}
where the wave function, wave function renormalization, and the
regular part of the Green's function have all become functions of the
constant scalar field $\sigma$, while $\epsilon_\sigma({\bf
p})=\sqrt{M(\sigma)^2 + {\bf p}^2}$.  Next, we insert
Eq.~(\ref{eq:GFsigma}) into Eq.~(\ref{eq:vertid}) and expand both
sides of the result in powers of $p_0 - \epsilon({\bf p})$. The terms
with a double pole at $p_0=\epsilon({\bf p})$ yield the identity
\begin{equation}
Z_2 \bar{u}({\bf p}) \Lambda(\overline{p},\overline{p}) u({\bf p})
=S~\left.\frac{d \epsilon_\sigma({\bf p})}{d \sigma}\right|_{\sigma=0}
- V. \label{eq:LSV1st}
\end{equation}
This agrees with the on-shell vertices obtained by Birse~\cite{Bi95}.
Meanwhile, the terms with a single pole at $p_0=\epsilon({\bf p})$ give
\begin{eqnarray}
Z_2 \bar{u}({\bf p}) \left. \frac{d\Lambda(p,p)}{d p_0} 
\right|_{p_0=\epsilon({\bf p})} u({\bf p})
+ u^\dagger({\bf p}) \delta g(\overline{p}) \Lambda(\overline{p},
\overline{p}) u({\bf p})
+ \bar{u}({\bf p}) \Lambda(\overline{p},
\overline{p}) \delta g(\overline{p}) \gamma_0 u({\bf p})
&=&\left.\frac{D Z_2(\sigma)}{Z_2}\right|_{\sigma=0}\nonumber\\
&=&S \left.\frac{d Z_2(\sigma)}{d \sigma} \right|_{\sigma=0}
\label{eq:identity},
\end{eqnarray}
where the last step follows since, 
as observed above, $Z_2$ does not depend upon the constant
vector field. The identities (\ref{eq:LSV1st}) and (\ref{eq:identity}) 
will be crucial in
our evaluation of $\bar{u}({\bf p}_f) {\cal A}(p_f,Q,0,p_i) u({\bf p}_i)$.

We are now ready to perform this evaluation. Consider the quantity 
\begin{eqnarray}
X({\bf p}_f,{\bf p}_i) &\equiv& 
\lim_{p_f^0 \rightarrow \epsilon({\bf p}_f)} \, \,
\lim_{p_i^0 \rightarrow \epsilon({\bf p}_i)}
u^\dagger({\bf p}_f) \frac{1}{2~Z_2} (p_f^0 - \epsilon({\bf p}_f)) G(p_f)
\nonumber\\
&\times& \biggl[\Lambda(p_f,p_i) G(p_i) \Lambda(p_i,p_i)
 + \Lambda(p_f,p_f) G(p_f) \Lambda(p_f,p_i)
+ C(p_f,Q,0,p_i) \biggr]\nonumber\\
&\times& G(p_i) (p_i^0 - \epsilon({\bf p}_i)) \gamma_0 u({\bf p}_i).
\label{eq:Xdef}
\end{eqnarray}
This form is motivated by LSZ reduction~\cite{Le55}, but it is not
exactly equal to the Feynman diagram expression (\ref{eq:intamp})
because in two of the three terms here there is a
zero-momentum-transfer vertex. The consequent presence of two poles at
the same point in $p_0$ means that Eq.~(\ref{eq:Xdef}) yields a
different result from Eq.~(\ref{eq:intamp}).  In order to develop the
relation of the vertices which appear in Eq.~(\ref{eq:intamp}) to the
zero-momentum-transfer vertices that appear in $X$, it is necessary to
expand as follows,
\begin{equation}
\Lambda (\bar{p},p) = \Lambda (p,p) + \frac{\partial \Lambda
(p',p)}{\partial p'_0} \biggr|_{p' = p} ( \epsilon ({\bf
p})  - p _0 ) + \cdots .
\label{eq:Lamexp}
\end{equation}
Using Eq.~(\ref{eq:Lamexp}) in order to expand the quantity
$(p^0  - \epsilon({\bf p})) G(p) \Lambda(p,p) G(p)$
in powers of $(p_0 - \epsilon({\bf p}))$ leads to 
\begin{eqnarray}
(p^0  - \epsilon({\bf p}))
G(p) \Lambda(p,p) G(p)=
Z_2 u({\bf p}) \bar{u} ({\bf p}) \Lambda(\overline{p},p) G(p) +
Z_2 u({\bf p}) \bar{u}({\bf p}) \left. 
\frac{\partial \Lambda(p',\overline{p})}{\partial p_0'}
\right|_{p'=\overline{p}} Z_2 u({\bf p}) \bar{u}({\bf p})
\nonumber\\
+ \delta g(\overline{p}) \Lambda(\overline{p},\overline{p})Z_2 u({\bf p}) 
\bar{u}({\bf p}) + {\cal O}[p_0 - \epsilon({\bf p})].
\label{eq:expansion}
\end{eqnarray}
From this result and the analogous expression for $G(p) \Lambda(p,p)
G(p) (p_0 - \epsilon({\bf p}))$ we find
\begin{equation}
X({\bf p}_f,{\bf p}_i)=\bar{u}({\bf p}_f)
{\cal A}({\overline{p_f},Q,0,\overline{p_i}) u({\bf p}_i)
+ {1 \over 2} Z_2 \bar{u}({\bf p}_f) \Lambda(\overline{p_f},\overline{p_i}) 
u({\bf p}_i) [f({\bf p}_f) + \tilde{f}({\bf p}_i)].}
\end{equation}
where
\begin{equation}
f({\bf p})= \bar{u}({\bf p}) \left.
\frac{\partial \Lambda(p',\overline{p})}{\partial p_0'}
\right|_{p'=\overline{p}} Z_2 u({\bf p})
+ u^\dagger({\bf p}) \delta g(\overline{p}) \Lambda(\overline{p},
\overline{p})u({\bf p})
\end{equation}
and 
\begin{equation}
\tilde{f}({\bf p})= \bar{u}({\bf p}) \left.
\frac{\partial \Lambda(\overline{p},p')}{\partial p_0'}
\right|_{p'=\overline{p}} Z_2 u({\bf p})
+ \bar{u}({\bf p}) \Lambda(\overline{p},\overline{p}) \delta g 
(\overline{p}) \gamma_0 u({\bf p}).
\end{equation}
The result (\ref{eq:identity}) may now be rewritten as
\begin{equation}
f({\bf p}) + \tilde{f}({\bf p})=\left. \frac{DZ_2 (\sigma)}{Z_2} 
\right|_{\sigma=0}.
\end{equation}
It is sufficient for the factor $f + \tilde{f}$ to be evaluated at $Q
= 0$ in order for the scalar part of $X({\bf p}_f, {\bf p}_i)$ to be
correct to zeroth order in $Q$ and the spin-dependent part to be
correct to first-order in $Q$.  Thus, if we are only concerned with
the spin-independent forward scattering and the first-order spin-orbit
interaction:
\begin{eqnarray}
\bar{u}({\bf p}_f){\cal A}(\overline{p_f},Q,0,\overline{p_i}) u({\bf p}_i)
=X({\bf p}_f,{\bf p}_i) 
+ {1 \over 2}Z_2 \left. D\left(\frac{1}{Z_2(\sigma)}\right)\right|_{\sigma=0}
Z_2 \bar{u}({\bf p}_f) \Lambda(\overline{p_f},\overline{p_i}) 
u({\bf p}_i).
\label{eq:Xreln}
\end{eqnarray}

Finally, note that Eqs.~(\ref{eq:contactdef}) and (\ref{eq:functid})
imply that the contact term where one leg carries zero
momentum transfer may be calculated via
\begin{equation}
C(p+Q/2,Q,0,p-Q/2)=\left.D 
\Lambda_\sigma(p+Q/2,p-Q/2)\right|_{\sigma=0}.
\label{eq:contactid}
\end{equation}
Here $\Lambda_\sigma$ is defined to be the vertex calculated 
in the presence of a constant $\sigma$ field. Using Eqs.~(\ref{eq:Xreln}), 
(\ref{eq:Xdef}), (\ref{eq:vertid}), and (\ref{eq:contactid}) then 
produces our central result:
\begin{eqnarray}
&&\bar{u}({\bf p}_f){\cal A}(\overline{p_f},Q,0,\overline{p_i}) 
u({\bf p}_i)
\nonumber\\
&& \qquad =
\lim_{p_f^0 \rightarrow \epsilon({\bf p}_f)}
\lim_{p_i^0 \rightarrow \epsilon({\bf p}_i)}
u^\dagger({\bf p}_f) \frac{1}{2}
(p_f^0 - \epsilon({\bf p}_f))
D [G_\sigma(p_f) \Lambda_\sigma(p_f,p_i) G_\sigma(p_i)/
Z_2(\sigma)]_{\sigma=0} \, \,
(p_i^0 - \epsilon({\bf p}_i)) \gamma_0 u({\bf p}_i),
\label{eq:crucial}
\end{eqnarray}
where $D$ acting on $G_\sigma$ generates the zero-momentum vertices
in (\ref{eq:crucial}), and $D$ acting on $\Lambda_\sigma$ produces
the contact term, as shown in Eq.~(\ref{eq:contactid}).
The division by $Z_2(\sigma)$ arises because of the effect the presence of
zero momentum-transfer vertices has on the taking
of the on-shell limits. Physically, it is necessary in order
to ensure the correct normalization of the asymptotic states---a delicate
issue in a calculation involving a constant external field which 
alters $Z_2$.

We now deal separately with the scalar and vector pieces of the $D$
operator in Eq.~(\ref{eq:crucial}).  The vector piece, which arises
from the piece of $D$ which is proportional to $V$, may be simplified
using the decomposition (\ref{eq:GFsigma}) and Eq.~(\ref{eq:LSV1st}),
into the form
\begin{eqnarray}
V \left\{ \frac{1}{p_f^0 - \epsilon({\bf p}_f)}
{1 \over 2}\bar{u}({\bf p}_f) Z_2 \Lambda(p_f,p_i)
u({\bf p}_i)
+ {1 \over 2}\bar{u}({\bf p}_f) Z_2 \Lambda(p_f,p_i)
u({\bf p}_i) \frac{1}{p_i^0 - \epsilon({\bf p}_i)}
- {1 \over 2}\bar{u} ({\bf p}_f) Z_2 \frac{d\Lambda(p_f,p_i)}{d p_0} 
u({\bf p}_i) \right\}.
\label{eq:vectorbit}
\end{eqnarray}
Note that in deriving this expression the $\delta g$ pieces of the
decomposition (\ref{eq:GFsigma}) do not contribute in the limit 
$p_i \rightarrow \overline{p_i}$, $p_f \rightarrow \overline{p_f}$.
Expanding out the first two terms in powers of $p_f^0 - 
\epsilon({\bf p}_f)$ and $p_i^0 - \epsilon({\bf p}_i)$, respectively,
taking the limit as the external legs are put on shell, and using 
Eq.~(\ref{eq:LSV1st}) allows us to rewrite (\ref{eq:vectorbit}) as
\begin{eqnarray}
{1 \over 2} V_V^{(1)}({\bf p}_f,{\bf p}_f)
\frac{1}{p_f^0 - \epsilon({\bf p}_f)}
V^{(1)}({\bf p}_f,{\bf p}_i)
+ {1 \over 2} V^{(1)}({\bf p}_f,{\bf p}_i)
\frac{1}{p_i^0 - \epsilon({\bf p}_i)}
V_V^{(1)}({\bf p}_i,{\bf p}_i),
\end{eqnarray}
where
\begin{equation}
V^{(1)}({\bf p}',{\bf p})=Z_2 \bar{u}({\bf p}') 
\Lambda(\overline{p}',\overline{p})u({\bf p}),
\end{equation}
and $V_V^{(1)}$ is the piece of $V^{(1)}$ which is proportional to
$V$.

As for the scalar piece of the second-order interaction  ({\it i.e.}
the result generated by the action of the $S \frac{\partial}{\partial
\sigma}$ in Eq.~(\ref{eq:crucial})), inserting the expression 
(\ref{eq:GFsigma}) in Eq.~(\ref{eq:crucial}) and then expanding about
the on-shell points, taking the appropriate limits, and using
Eq.~(\ref{eq:LSV1st}) and the identity
\begin{equation}
u^\dagger({\bf p})\left.\frac{du_\sigma({\bf p})}{d\sigma}\right|_{\sigma=0}=0,
\end{equation}
leads to
\begin{eqnarray}
{1 \over 2}V_S^{(1)}({\bf p}_f,{\bf p}_f)
\frac{1}{p_f^0 - \epsilon({\bf p}_f)}
V^{(1)}({\bf p}_f,{\bf p}_i)
+ {1 \over 2}V^{(1)}({\bf p}_f,{\bf p}_i)
\frac{1}{p_i^0 - \epsilon({\bf p}_i)}
V_S^{(1)}({\bf p}_i,{\bf p}_i)
\nonumber\\
+ {1 \over 2} S  \frac{d}{d \sigma} [\bar{u}_\sigma({\bf p}_f) Z_2(\sigma) 
\left.\Lambda_\sigma(p_f,p_i) \right|_{p_0=\epsilon_{\sigma}({\bf p})}
u_\sigma({\bf p}_i)]_{\sigma=0},
\label{eq:scalarbit}
\end{eqnarray}
where $V_S^{(1)}$ is the piece of $V^{(1)}$ which is proportional to
$S$.

Observe that the sum of Eqs.~(\ref{eq:vectorbit}) and (\ref{eq:scalarbit}) 
contains the iterate of the on-shell 
first-order interaction, $V^{(1)}$. 
In Ref.~\cite{Wa95} it was shown that if amplitudes such
as ${\cal V}^{(2)}$ were to be used 
as a potential in a wave equation then the iterated on-shell
first-order interaction had to be subtracted. Once this divergent piece of
the interaction ${\cal V}^{(2)}$ is removed and Eq.~(\ref{eq:potform}) invoked
we have the following general result 
for the potential:
\begin{eqnarray}
&&\overline{{\cal V}}^{(2)}({\bf p} + {\bf Q}/2,{\bf p} - {\bf Q}/2)=
\nonumber\\
&& \quad \frac{1}{2} \int \frac{ d^3q}{(2 \pi)^3} \, h({\bf Q}/2 + {\bf q}) 
h({\bf Q}/2 - {\bf q}) S
\frac{d}{d \sigma} [\bar{u}_\sigma({\bf p}+{\bf Q}/2) Z_2(\sigma) \left.
\Lambda_\sigma(p+Q/2,p-Q/2) \right|_{p_0=\epsilon_{\sigma}({\bf p})}
u_\sigma({\bf p}-{\bf Q}/2)]_{\sigma=0}.
\label{eq:answer}
\end{eqnarray}
Here the value of $Q_0$ is understood to be $\epsilon({\bf p} + {\bf
Q}/2) - \epsilon({\bf p} - {\bf Q}/2)$.  The spin-independent part of
(\ref{eq:answer}) is correct to zeroth order in ${\bf Q}$ and the
spin-dependent part is correct to first-order in ${\bf Q}$.  We
observe that at least one interaction here must be with the scalar
field because the vector-vector piece of the irreducible potential is
zero.

The result (\ref{eq:answer}) is extremely general. It relies only
on the assumption that the vector field couples to a conserved current.
{\it No assumptions about the structure of the composite or its couplings to
the scalar field have been made.} 

Note that in the limit of constant fields ($h({\bf Q})=\delta^{(3)}(Q)$), 
we may use Eq.~(\ref{eq:LSV1st}) in Eq.~(\ref{eq:answer}) and so reproduce 
the result of Birse~\cite{Bi95}:
\begin{eqnarray} 
\overline{{\cal V}}^{(2)}({\bf p}+{\bf Q}/2,{\bf p}-{\bf Q}/2)&=&
\delta^{(3)}(Q) \frac{S^2}{2}
\left. \frac{d^2 \epsilon_{\sigma}({\bf p})}{d \sigma^2} \right|_{\sigma=0}
\nonumber \\
&=& \left\{ \frac{S^2}{2} \frac{M}{\epsilon({\bf p})} 
\frac{d^2M}{d \sigma ^2} + 
\frac{S^2 {\bf p}^2}{2 \epsilon ^3 ({\bf p})} \left( \frac{dM}{d \sigma} 
\right)^2 \right\} 
\delta ^{(3)} ({\bf Q}) . 
\label{eq:fwdscat}
\end{eqnarray} 
Hence Eq.~(\ref{eq:answer}) encompasses the result obtained in the
model of Refs.~\cite{Wa94,Wa95} when $M$ is a linear function of
the scalar interaction.
More generally, the forward scattering
amplitude is found to agree with that predicted by the Dirac
equation when the mass is a function of the scalar interaction,
$M=M(\sigma(r))$.  The spin-independent interaction at second
order has two components; one involving the second derivative of
$M(\sigma)$, which is a quantity that depends on the internal
structure of the composite system, and the other involving the
square of the first-order interaction, $\frac{dM}{d \sigma} S$,
times a model-independent kinematical factor.  

In the analysis of Refs. \cite{Wa94,Wa95}, contributions from
off-shell vertices, off-shell propagation of the intermediate state
and the contact term, all of which must be considered for a composite
particle, cancel. Consequently, the term proportional to ${\bf p}^2$,
which in that model arises from composite particle z-graphs, is the
complete result for $\overline{{\cal V}}^{(2)}$ at zero momentum
transfer. In the approach of this paper Eq.~(\ref{eq:answer}) must be
applied in order to calculate $\overline{{\cal V}}^{(2)}$ in the limit
${\bf Q} \rightarrow 0$. When this is done a number of terms are
generated. The derivative acting on $\Lambda_\sigma$ generates the
contact interaction. Off-shell pieces of the vertices arise from $D$'s
action on both the constraint $p_0=\epsilon_\sigma({\bf p})$ and
$Z_2(\sigma)$. Furthermore, Eq.~(\ref{eq:identity}) shows that
$DZ_2(\sigma)$ also generates contributions from off-shell
intermediate-state propagation. Consequently, all the pieces that
cancel in the analysis of Refs.~\cite{Wa94,Wa95} are generated in the
more general result (\ref{eq:answer}). However, in contrast to
Refs.~\cite{Wa94,Wa95}, here we see that they add to give
the first term in Eq.~(\ref{eq:fwdscat}). In the specific model of
Refs.~\cite{Wa94,Wa95} this term was identically zero.  Finally, the
second term in Eq.~(\ref{eq:fwdscat}) may be shown to arise from the
parts of Eq.~(\ref{eq:answer}) involving derivatives of the wave
function $u_\sigma({\bf p})$ with respect to $\sigma$.  Such a
derivative gives a term involving the negative-energy wave function
$v(-{\bf p})$, i.e.,
\begin{equation}
\frac{d u_{\sigma} ({\bf p})}{d \sigma} \biggr|_{\sigma = 0} =
\frac{v(-{\bf p}) \bar{v}(-{\bf p})}{2 \epsilon ({\bf p})} \frac{dM}{d
\sigma} u({\bf p}).   
\label{eq:dudsigma}
\end{equation}

\section{Calculation of the spin-orbit interaction}

\label{sec-five}

In this section we calculate the actual value of the irreducible piece of the
second-order spin-orbit potential, 
$\overline{{\cal V}}^{(2)}({\bf p}+{\bf Q}/2,{\bf p}-{\bf Q}/2)$.
In order to use Eq.~(\ref{eq:answer}), the general form of the 
first-order vertex $\Lambda_\sigma(p+Q/2,p-Q/2)$ 
must be written down.
In the scalar case, up to first-order in $Q$
the most general function $\Lambda^S$ which is consistent with
Lorentz invariance, is
\begin{equation}
\Lambda^S(p+Q/2,p-Q/2)= S\bigl\{ 
a_S + b_S(p^2) \not \! p + c_S(p^2) [\not \! p,\not \! Q]
+ d_S(p^2) p \cdot Q \bigr\}.
\end{equation}
However, as discussed in Sec.~~\ref{sec-two}, $u({\bf p})$ is proportional
to a Dirac spinor corresponding to mass $M$. Thus
\begin{eqnarray}
\bar{u}({\bf p} + {\bf Q}/2) \, p \cdot Q \, u({\bf p} - {\bf Q}/2)&=&0\\
\bar{u}({\bf p} + {\bf Q}/2) \, [\not \! p,\not \! Q] \, 
u({\bf p} - {\bf Q}/2)&=& {\cal O}(Q^2).
\end{eqnarray}
Therefore, to first-order in ${\bf Q}$ we have
\begin{equation}
V_S^{(1)}({\bf p} + {\bf Q}/2,{\bf p} - {\bf Q}/2)= S
\left[ a_S + b_S(M^2) M \right] \bar{u}({\bf p} + {\bf Q}/2) u({\bf p} 
- {\bf Q}/2).
\end{equation}
By going to the limit ${\bf Q}=0$ and comparing with (\ref{eq:LSV1st})
we identify $a_S + b_S(M^2)=\frac{dM}{d \sigma}$, thus showing that to first
order in ${\bf Q}$ the spin-orbit piece of the scalar potential is
\begin{equation}
V_S^{(1)}({\bf p} + {\bf Q}/2,{\bf p} - {\bf Q}/2)
= \frac{dM}{d \sigma} 
\frac{i}{2 \epsilon({\bf p})}S \frac{1}{\epsilon({\bf p}) + M} {\bf \sigma}
\cdot ({\bf p} \times {\bf Q}).
\end{equation}

Furthermore, the most general form 
for the vector vertex, consistent with the use of minimal substitution
as in (\ref{eq:minsub}), is:
\begin{equation}
V_V^{(1)}({\bf p} + {\bf Q}/2,{\bf p} - {\bf Q}/2)
= \bar{u}({\bf p} + {\bf Q}/2) V [F_1(Q^2) \gamma_0
+ F_2 (Q^2) \frac{i}{2M} \sigma^{0 \mu} Q_\mu] u({\bf p}
- {\bf Q}/2),
\end{equation}
with $F_1(0)=1$ and $F_2(0)=\kappa$, the anomalous magnetic moment. Thus, 
to first-order in ${\bf Q}$ the most general form of the total first-order
spin-orbit potential is
\begin{equation}
V^{(1)}_{SO}({\bf p}+{\bf Q}/2,{\bf p}-{\bf Q}/2)=
\left(\frac{d M}{d \sigma} S - V - 
V \kappa \frac{\epsilon({\bf p}) + M}{2 M}\right)
\frac{i}{2 \epsilon({\bf p})}\frac{1}{\epsilon({\bf p}) + M} {\bf \sigma}
\cdot ({\bf p} \times {\bf Q}).
\label{eq:genform}
\end{equation}
In the presence of a scalar field the mass $M$, and hence $\epsilon$, 
as well as the anomalous magnetic moment $\kappa$, all become
dependent on $\sigma$. So, from (\ref{eq:genform}) and (\ref{eq:answer}) we 
derive:
\begin{eqnarray}
&& \overline{{\cal V}}_{SO}^{(2)}({\bf p},{\bf Q})=
\frac{1}{2} \int d^3q \, \left[h({\bf Q}/2 + {\bf q}) 
h({\bf Q}/2 - {\bf q})\right] \times
\label{eq:conclusion}\\
&& \qquad \left[-(S \frac{dM}{d \sigma} - V) S \frac{dM}{d\sigma}
\frac{1}{2 \epsilon^3({\bf p})} +
S^2 \frac{d^2 M}{d \sigma^2}
\frac{1}{\epsilon({\bf p}) + M} \frac{1}{2 \epsilon({\bf p})}
+ SV \frac{dM}{d \sigma} \kappa 
\frac{\epsilon^2({\bf p}) + M^2}{4 \epsilon^3({\bf p}) M^2}
- SV \frac{d \kappa}{d \sigma} \frac{1}{4 \epsilon({\bf p}) M} \right]
i {\bf \sigma} \cdot ({\bf p} \times {\bf Q}),
\nonumber
\end{eqnarray}
where all quantities are understood to be evaluated at $\sigma=0$.

\section{Conclusion}

\label{sec-conc}

As claimed in the Introduction, both the spin-independent and
spin-dependent second-order interactions agree with those obtained
from the Dirac equation at low momentum transfer, provided that the
scalar potential is introduced to the Dirac equation through a
replacement of the mass by $M(\sigma(r))$, and the anomalous magnetic
moment also is made a function of the scalar potential,
$\kappa(\sigma(r))$. This result must hold for any composite-particle
model and any scalar interaction.  The vector interaction is limited
to one which couples to a conserved charge of the composite system.
The choice of the interpolating field made in Eq.~(\ref{eq:Gdef}) is,
apart from its transformation properties, entirely
arbitrary and should have no effect on the results because amplitudes
are examined only at the composite-particle pole.

This paper suggests that when the compositeness of the
nucleon is considered, two differences from the practices followed in
Dirac phenomenology for nucleon-nucleus scattering appear. First, the
introduction of the scalar interaction though a function $M(\sigma(r))$
implies that nonlinear terms such as ${1 \over 2} \frac{\partial^2
M}{\partial \sigma ^2}$ arise. Such terms depend upon the internal
structure of the nucleon and are model dependent. Similar nonlinear
terms are implied in the interaction between two free
nucleons. However, introducing such terms into the Dirac phenomenology
is not entirely straightforward.  Since the impulse approximation is
used to construct the Dirac potentials from the empirically determined
$NN$ interaction, the contribution this nonlinear effect makes to
the two-body interaction is already included in Dirac phenomenology
calculations.  However, the nonlinear terms also imply three-body and
higher interactions and these are {\it not} usually included.  Second, while
the anomalous magnetic moment typically {\it is} included in Dirac
phenomenology, the dependence of $\kappa$ on $\sigma$ is model dependent,
and is generally omitted.

\acknowledgements{We thank J.~A. McGovern for useful conversations.
D.~R.~P. thanks J.~A. Tjon for challenging him to think
further about the meaning of Eq.~(\ref{eq:crucial}). We are grateful to
the U.S. Department of Energy for its support under contract no. 
DE-FG02-93ER-40762. M.~C.~B. acknowledges support from the EPSRC.}


\begin{thebibliography}{10}

\bibitem{Bi95}
M.~C. Birse, Phys. Rev. C {\bf 51},  1083  (1995).

\bibitem{Wa94}
S.~J. Wallace, F. Gross, and J.~A. Tjon, Phys. Rev. Lett. {\bf 74},  228
  (1995).

\bibitem{Wa95}
S.~J. Wallace, F. Gross, and J.~A. Tjon, Phys. Rev. C {\bf 53},  860  (1996).

\bibitem{Wa87}
S.~J. Wallace, Ann.\ Rev.\ Nucl.\ Part.\ Sci.\ {\bf 37},  267  (1987).

\bibitem{Br84}
S.~J. Brodsky, Comm. Nucl. Part. Phys. {\bf 12},  213  (1984).

\bibitem{JB91}
T. Jaroszewicz and S.~J. Brodsky, Phys. Rev. C {\bf 43},  1946  (1991).

\bibitem{Co92}
T.~D. Cohen, Phys. Rev. C {\bf 45},  833  (1992).

\bibitem{AW88}
J. Achtzehnter and L. Wilets, Phys. Rev. C {\bf 38},  5  (1988).

\bibitem{Lo54}
F.~E. Low, Phys. Rev. {\bf 96},  1428  (1954).

\bibitem{GG54}
M. Gell-Mann and M.~L. Goldberger, Phys. Rev. {\bf 96},  1433  (1954).

\bibitem{Io81}
B.~L. Ioffe, Nucl. Phys. {\bf B188},  317  (1981).

\bibitem{IZ84}
C. Itzykson and J. Zuber, {\em Quantum Field Theory} (McGraw-Hill, Singapore,
  1985).

\bibitem{Le55}
H. Lehmannn, K. Symanzik, and W. Zimmerman, Nuovo Cim. {\bf 1},  205  (1955).

\end{thebibliography}
\end{document}